\documentclass[letterpaper, 10 pt, conference]{ieeeconf}  
    \usepackage{graphicx}
\usepackage{mathtools} 
\usepackage{amsmath}
\usepackage{amsfonts}
\usepackage{amssymb}
\usepackage{dsfont}
\usepackage{tikz}
\usepackage{lipsum}
\usepackage{multicol}
\usepackage{graphicx,graphics}
\usepackage{dirtytalk}
\usepackage{algorithm}
\usepackage{algpseudocode}
\usepackage{multirow,array}
\usepackage{xcolor}
\usepackage{color,soul}
\makeatletter
\newcommand\fs@betterruled{%
  \def\@fs@cfont{\bfseries}\let\@fs@capt\floatc@ruled
  \def\@fs@pre{\vspace*{5pt}\hrule height.8pt depth0pt \kern2pt}%
  \def\@fs@post{\kern2pt\hrule\relax}%
  \def\@fs@mid{\kern2pt\hrule\kern2pt}%
  \let\@fs@iftopcapt\iftrue}
\floatstyle{betterruled}
\restylefloat{algorithm}
\makeatother
\usepackage{tikz}
\usetikzlibrary{arrows}
\usetikzlibrary{shapes}
\usetikzlibrary{shapes.geometric}
\usetikzlibrary{automata, arrows.meta, positioning}
\tikzset{
  invisible/.style={opacity=0},
  dots/.style={state,draw=none},
  visible on/.style={alt={#1{}{invisible}}},
   alt/.code args={<#1>#2#3}{%
     \alt<#1>{\pgfkeysalso{#2}}{\pgfkeysalso{#3}} 
  },
}
\usepackage{textcomp}                                     

\IEEEoverridecommandlockouts                              
\title{\LARGE \bf
On Bellman's principle of optimality and Reinforcement learning for safety-constrained Markov decision process
}

\newtheorem{lemma}{Lemma}
\newtheorem{definition}{Definition}

\newtheorem{assumption}{Assumption}

\newcommand\norm[1]{\left\lVert#1\right\rVert}

\author{Rahul Misra$^{*}$, Rafa\l \hspace{0.01cm} Wisniewski and Carsten Skovmose Kallesøe
\thanks{*This work was supported by Poul Due Jensens Fonden project SWIfT}
\thanks{R. Misra, R. Wisniewski and C. Kallesøe are with the Department of Electronic Systems, Automation and Control, Aalborg
University, Fredrik Bajers Vej 7 C, 9220 Aalborg East, Denmark
        {\tt\small rmi@es.aau.dk, raf@es.aau.dk, csk@es.aau.dk}}%
}

\begin{document}

\maketitle
\thispagestyle{empty}
\pagestyle{empty}

\begin{abstract}
We study optimality for the safety-constrained Markov decision process which is the underlying framework for safe reinforcement learning. Specifically, we consider a constrained Markov decision process (with finite states and finite actions) where the goal of the decision maker is to reach a target set while avoiding an unsafe set(s) with certain probabilistic guarantees. Therefore the underlying Markov chain for any control policy will be multichain since by definition there exists a target set and an unsafe set. The decision maker also has to be optimal (with respect to a cost function) while navigating to the target set. This gives rise to a multi-objective optimization problem. We highlight the fact that Bellman's principle of optimality \cite{bellman1957dynamic} may not hold for constrained Markov decision problems with an underlying multichain structure (as shown by the counterexample in \cite{haviv1996constrained}). We resolve the counterexample in \cite{haviv1996constrained} by formulating the aforementioned multi-objective optimization problem as a zero-sum game and thereafter construct an asynchronous value iteration scheme for the Lagrangian (similar to Shapley's algorithm \cite{shapley1953stochastic}). Finally, we consider the reinforcement learning problem for the same and construct a modified $Q$-learning algorithm for learning the Lagrangian from data. We also provide a lower bound on the number of iterations required for learning the Lagrangian and corresponding error bounds.     
\end{abstract}

\section{Introduction}
 
In recent years, due to the success of Reinforcement learning (RL) techniques, there has been a renewed interest in the study of Markov decision processes (MDP) which is the underlying mathematical framework for RL. A number of applications of RL are safety-critical (such as robotics \cite{kober2013reinforcement}, autonomous cars \cite{lutjens2019safe} and healthcare \cite{tejedor2020reinforcement}) and this motivates research on safe RL. This work combines literature from three distinct research fields which are: Reach-avoid problems, Constrained MDPs, and RL for Constrained MDPs. 
\subsection{Reach-Avoid problems}
A classical problem in control theory is to ensure that the state trajectory reaches a target set while avoiding an unsafe set. This is referred to as \textit{Reach-Avoid} problem in \cite{ding2010robust}, \cite{summers2011stochastic} and \cite{esfahani2016stochastic} and references therein. Many variants of this problem (in different settings) have been addressed in the literature. A Robust Reach-Avoid problem is solved for switched systems in \cite{ding2010robust}, while \cite{summers2011stochastic} considers the Reach-Avoid problem for Discrete-Time Stochastic Hybrid systems with time-varying unsafe sets. The Reach-Avoid problem for systems with dynamics defined via stochastic differential equations was solved in \cite{esfahani2016stochastic}. References \cite{wisniewski2021safe} and \cite{bujorianu2020stochastic} consider stochastic safety for MDPs and our work builds up on the results of \cite{wisniewski2021safe}. To the best knowledge of authors, there is a shortage of literature addressing the Reach-Avoid problem for MDPs and specifically, in RL setting with unknown MDPs. 
\subsection{Constrained MDP}
Constrained MDP problems have been studied to model multi-objective optimization problems in \cite{altman1999constrained} and safety-constrained MDPs in \cite{el2018controlled}. An important distinction made in the study of MDPs is whether the underlying Markov chain for a given policy is unichain (also referred to as ergodic Markov chain) or multichain \cite{puterman2014markov}. The celebrated Bellman's principle of optimality does not hold in general for constrained MDPs with underlying multichain structure as shown in the counterexample due to \cite{haviv1996constrained}. Attempts have been made to resolve this counterexample for average-reward MDPs in \cite{chong2012bellman} and for finite time-horizon MDPs in \cite{chow2015time}. Reference \cite{chong2012bellman} proposes changing constraints as new states are visited (i.e. constraint changes depending on previous outcomes). This is not entirely satisfactory in our case as safety guarantees should not be changed based on previous outcomes. Reference \cite{chow2015time} proposes dynamic programming formulation with time consistency for risk-constrained stochastic optimal control problems wherein they introduce an additional dynamic programming operator which ensures the selection of only feasible policies. This is a similar approach as in \cite{chen2004dynamic} and is not satisfactory for us since it requires the computation of a feasible set at each time step in addition to the standard dynamic programming operator. Therefore, this approach is computationally more expensive with an increasing number of control actions, states as well as constraints (higher 'curse of dimensionality' compared to the standard dynamic programming).
\subsection{RL for constrained MDPs}
Despite the aforementioned counterexample, there are many works on RL for constrained MDPs such as an actor-critic algorithm proposed in \cite{borkar2005actor}, the natural policy gradient primal-dual proposed in \cite{ding2020natural} and exploration-exploitation bounds for constrained MDPs is addressed in \cite{efroni2020exploration}. This is because it is a common assumption that the underlying Markov chain has a unichain structure and the counterexample is for an MDP with an underlying multichain structure. The paper \cite{mannor2004geometric} is the closest related to our work as it considers the constrained MDP problem to be a $2$-player zero-sum game where one player is the agent and is trying to reach a target set and the other player is an adversary (nature or other adversarial agents). Learning for the average-reward case is analyzed and Blackwell's Approachability Theorem is used to prove convergence to the target set. However, the considered MDP is still an unichain MDP in \cite{mannor2004geometric}.

In this work, we consider the safe RL problem as a constrained multichain MDP problem and present algorithms for the same. The contributions are summarized as follows.
\begin{itemize}
    \item Bellman's Principle of Optimality is studied for constrained MDPs with stopping time.
    \item An algorithm is proposed for finding policies (which satisfy Bellman's principle of Optimality) for the counterexample due to \cite{haviv1996constrained}.
    \item An off-policy modified $Q$-learning algorithm is proposed for learning policies (which satisfy Bellman's principle of Optimality) with finite-time error bounds. 
\end{itemize}

The rest of the paper is organized as follows: Section II summarizes some common notation used throughout this paper. Section III formally introduces the problem formulation. Bellman's principle of Optimality is formally stated and the counterexample due to \cite{haviv1996constrained} is introduced in Section IV. An algorithm for asynchronous computation of Lagrangian (in the spirit of asynchronous value iteration) is proposed in Section V. In Section VI, we consider the problem of RL for constrained MDPs and propose an off-policy algorithm for learning the Lagrangian up to some error bounds. Finally, the conclusion and future work is presented in Section VII.  

\section{Notation}
\begin{itemize}
    \item $\mathds{1}_{(\cdot)}$ denotes a vector of $1$'s with appropriate dimension $(\cdot)$ indicated in the subscript.
    \item $\mathds{E}_{(\cdot)}$ represents the expectation operator corresponding to the probability measure $(\cdot)$ indicated in the subscript.
    \item $\Delta(A)$ represents a probability simplex in $\mathds{R}^{\mid A \mid}$ where $A$ is a finite set.
    \item $\mathcal{I}(a)$ represents the indicator function with 
    \begin{equation}
        \mathcal{I}(a) = \begin{cases}0, & \text {if } a \leq 0, \\ \infty, & \text { otherwise. }\end{cases}
    \end{equation}
\end{itemize}  
\section{Problem Formulation}
\begin{figure}[h] 
    \centering
    \includegraphics[width=8.4cm]{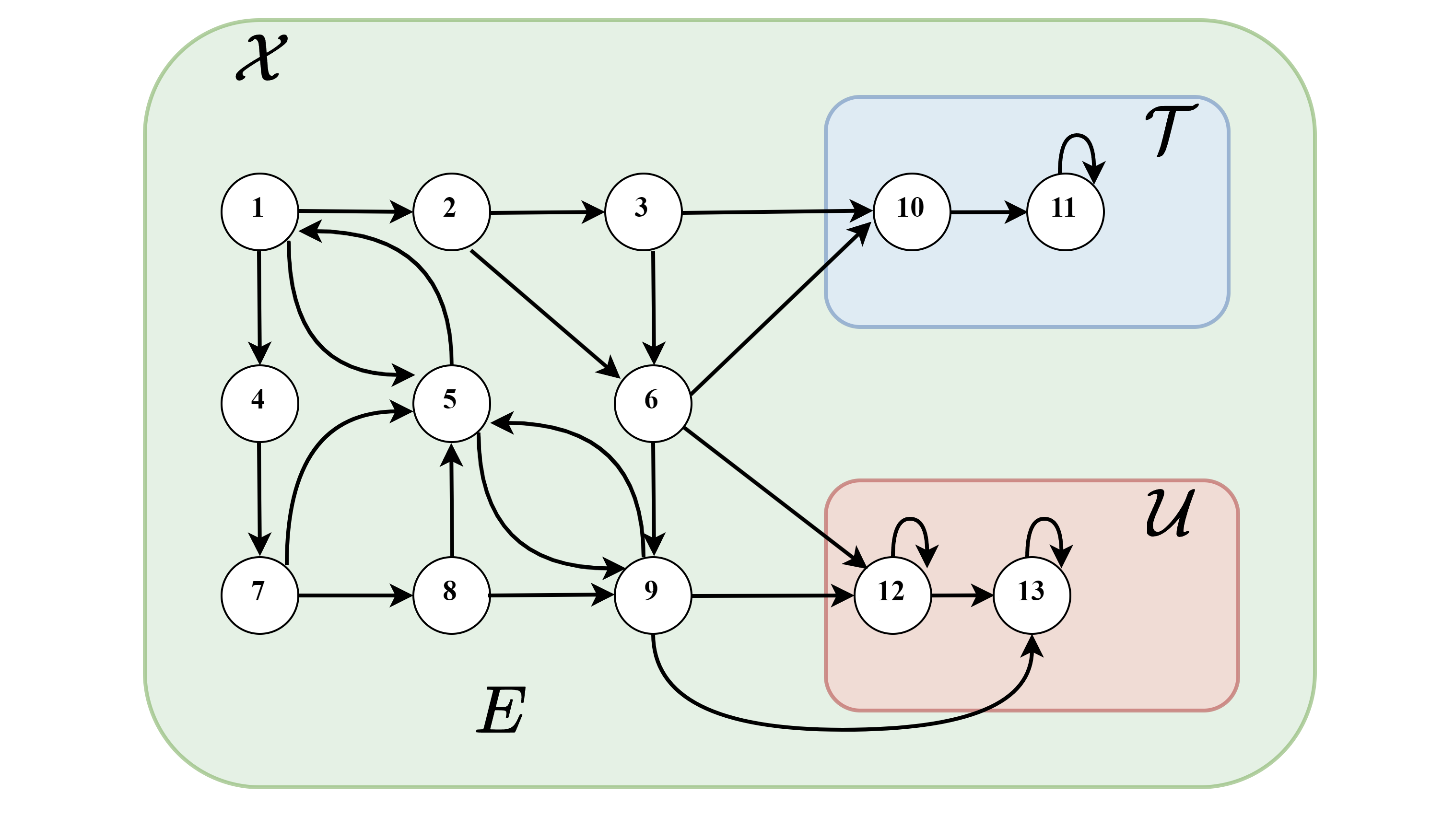}
    \caption{A generic Reach-Avoid problem for MDP with states numbered $1,\cdots, 13$. $\mathcal{X}$ is the entire state space partitioned into the target set $\mathcal{T}$, unsafe set $\mathcal{U}$ and transient states $E$.}
    \label{fig:Reach-avoid}
\end{figure}

We consider the reach-avoid problem for a Markov decision process where the goal of the decision maker is to pick control actions such that the state reaches a target set $\mathcal{T}$ while simultaneously avoiding some unsafe set $\mathcal{U}$ (with a certain probabilistic guarantee). A natural way to formulate this problem is to consider a constrained Markov decision process. Consider a stochastic process $(X_t)$ defined on the state space $\mathcal{X}$. The state space is partitioned into three sets: a set of transient states $E$, the target set $\mathcal{T}$, and an unsafe set $\mathcal{U}$ (see fig. \ref{fig:Reach-avoid}). The process $X_t$ starts in the set of non-absorbing states denoted by $E$ (referred to as transient states as $X_t$ leaves $E$ in finite time). We consider multichain MDPs which are defined as follows.  
\begin{definition}{(Multichain MDP)}
    An MDP is said to be multichain if there exists a policy $\pi$ such that the corresponding Markov chain consists of at least two ergodic chains and a (possibly empty) set of transient states.  
\end{definition}
The constrained MDP considered in this work is a multichain MDP as it consists of at least two recurrent chains: target set $\mathcal{T}$ and unsafe set $\mathcal{U}$. Let $A$ be a set of a finite number of actions. For two functions $c,k:\mathcal{X} \times A \to \mathds{R}$ representing the cost and the safety cost (defined later) respectively, let $c_t:= c(X_t, A_t)$, and $k_t:= k(X_t, A_t)$, where $A_t$ is a stochastic process defined on the set of actions $A$ which represents the control action taken at time $t$. In this work, we restrict our attention to $X_t$ as it evolves on transient state space $E$.  This is a natural way to study the Reach-Avoid problem as the study of $X_t$ is no longer interesting to us once it reaches either the target set or the unsafe set. Therefore, the sets $\mathcal{T}$ and $\mathcal{U}$ consist only of absorbing states where $c_t = 0$ and $k_t = 0$. Since $X_t$ stops once it exits $E$, we need to introduce stopping time $\tau$ which quantifies the random exit time of $X_t$ from $E$.   
\begin{definition}{(Stopping time)}
    Let $\tau$ denote the first hitting time of process $X_t$ to either $\mathcal{T}$ or $\mathcal{U}$. Formally $\tau:= \{\inf_t \mid X_t \in \mathcal{T} \cup \mathcal{U}\}$. 
\end{definition}
The control policies considered in this work are state-feedback memoryless policies (i.e. Markov policies) which are defined as follows.
\begin{definition}{(Markov policies $\pi$)}
    A Markov policy $\pi:E \to \Delta(A)$ is a stationary (i.e. independent of time) control policy that is dependent only on the given state and is defined as 
    \begin{equation}\label{Eq.Policy}
        \pi(x) = [\pi(a_1\mid x), \cdots, \pi(a_{\mid A \mid}\mid x)]\in \Delta(A).  
    \end{equation}
\end{definition}
We state the safety-constrained MDP problem with a probabilistic safety guarantee given by $w$ as follows.
\begin{definition}(Safety-constrained MDP problem)
    For $0 \leq w \leq 1$, we want to compute the minimum $V^*(i)$ of the cost
    \begin{align}\label{Eq.Optimize}
        V_{\pi}(i) := \mathds E_{\pi} \left[\sum_{t=0}^{\tau} c_t| x_0 = i\right],\text{ } i \in E,
    \end{align} 
    subject to
    \begin{align}\label{Eq.Constraints}
        W_{\pi}(i) := \mathds E_{\pi} \left[\sum_{t=0}^{\tau} k_t| x_0 = i\right] \leq w\text{, } i \in E,  
    \end{align}
    where $x_t \in E$ is the state at time $t$.    
\end{definition}
We denote the vector value function and vector safety function as $\mathbf{V}_{\pi} = (V_\pi(i))_{i \in E} \in \mathbb{R}^N$ and $\mathbf{W}_{\pi} = (W_\pi(i))_{i \in E} \in \mathbb{R}^N$ respectively for a policy $\pi$. For each stationary randomized policy $\pi: E \to \Delta(A)$, let $P(\pi)$ denote the probability transition matrix for the resulting Markov chain. The entries of matrix $P(\pi)$ are calculated as
\begin{align} \label{eq:P_pi}
    p_{ij}(\pi) = \sum_{a \in A} p_{ij}(a) \pi(a \mid i)
\end{align} 
The matrix $P(\pi)$ will be sub-stochastic (i.e. its entries sum to $< 1$) as the process $X_t$ has a positive probability of stopping at any given time due to random stopping time $\tau$. Define the expected cost vector $C_{\pi}$ under policy $\pi$ whose $i^{th}$ component is written as follows
\begin{equation}
    C_{\pi}(i) := \mathds E_{\pi}[c(i,a)],
\end{equation}
and $P^\mathcal{U}_E(\pi)$ denote the probability matrix of reaching set $\mathcal{U}$ starting from set $E$ with policy $\pi$. The rows represent the transient states and the columns represent the unsafe states. Let $K_\pi := P_E^\mathcal{U}(\pi) \mathds{1}_{\mid\mathcal{U}\mid}$ represent the vector of probabilities of reaching the set $\mathcal{U}$ with components
\begin{equation} \label{eq:k_pi}
    K_\pi(i) = \sum_{a \in A} \sum_{j \in \mathcal{U}} p_{ij}(a)\pi(a \mid i).
\end{equation} 

\section{Bellman's Principle of Optimality}

In this section, we will state Bellman's principle of optimality for MDPs. This principle does not necessarily hold for constrained MDPs as will be shown later in this section. The principle of optimality for unconstrained MDPs as stated in \cite{bellman1957dynamic} is \textit{An optimal policy has the property that whatever the initial state and initial decision are, the remaining decisions must constitute an optimal policy with regard to the state resulting from the first decision}. Mathematically, it can be formulated as follows. 
Consider the unconstrained optimization problem defined by \eqref{Eq.Optimize} and \eqref{Eq.Policy}. We introduce the optimal Bellman operator for this problem as follows
\begin{equation}\label{eq:Bellman_Operator}
   \mathcal{B}_{\pi^*} [{V}_{\pi^*}(i)] = \min_{a \in A} \left[ c_t +  \sum^N_{j=1} p_{ij}(a) {V}_{\pi^*}(j) \right] \quad \forall i\in E,  
 \end{equation}
 where $\pi^*$ is an optimal Markovian policy. Let $\mathcal{B}^i_{\pi^*}$ denote the optimal Bellman operator if $x_0 = i$ and let $\mathcal{B}^j_{\pi^*}$ denote the optimal Bellman operator if $x_0 = j$ than $\pi^*$ should satisfy the following property
\begin{equation}\label{eq:Bellman_policy}
     \mathcal{B}^i_{\pi^*}[{\mathbf{V}}_{\pi^*}] = \mathcal{B}^j_{\pi^*}[{\mathbf{V}}_{\pi^*}] \quad \text{ for any } i,j \in E.
\end{equation}
Furthermore as noted in \cite{bellman1957dynamic}, \eqref{eq:Bellman_policy} indicates that the problem defined by \eqref{Eq.Optimize} and \eqref{Eq.Policy} has an optimal substructure and therefore, we can find the optimal policy $\pi^*$ by finding the fixed point of \eqref{eq:Bellman_Operator} for each state $i\in E$. 

However, the following counterexample due to \cite{haviv1996constrained} shows that \eqref{eq:Bellman_policy} is not true in general for constrained multichain MDPs (and therefore \eqref{eq:Bellman_Operator} does not hold for constrained multichain MDPs). Consider a constrained multichain MDP with three recurrent chains $1, 2, 3$, three unsafe sets $S_1$, $S_2$, and $S_3$, and two transient states $i,j \in E$. If the process reaches Chain $1$, then the probability of hitting the unsafe set $S_1$ is $P(S_1) = 0.2$. If the process does not hit $S_1$ despite being in Chain $1$ then we say it has reached the target set. Similar is the case for Chains $2$ and $3$ with unsafe sets $S_2$ and $S_3$. Note, that the process terminates once it reaches either of the recurrent chains and hits the unsafe sets or it reaches either of the recurrent chains and avoids hitting the unsafe sets. If the process starts at state $i$, then with $0.5$ probability, it will go either to Chain $1$ or to state $j$. The decision maker can only decide whether to choose action $a$ or action $b$. The goal of the decision-maker is to have the expected probability of visiting $S_1 \cup S_2 \cup S_3 \leq 0.125$.   
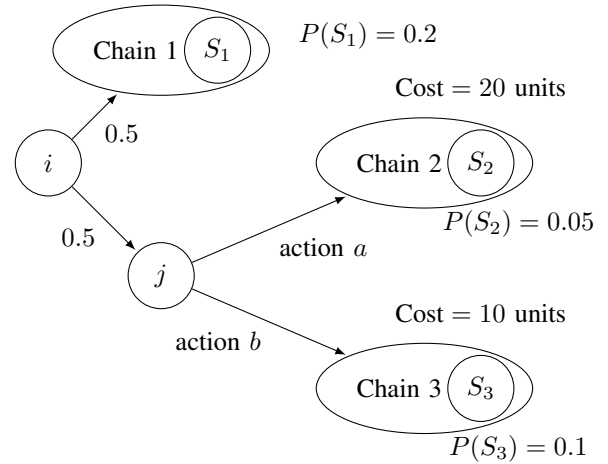
\begin{figure}[ht]
\centering
    \begin{tikzpicture} 
        \node[state] at (0,1.5) (x) {$i$};
        \node[ellipse, draw, align = left, text width = 1.8cm,
              minimum width = 2cm, 
              minimum height = 1.2cm] (c1) at (1.5,3) {Chain 1};
        \node[below] at (4.25,3.5) {$P(S_1) = 0.2$};
        \node[state] at (2.25,3) (s1) {$S_1$};
        \node[state] at (1.5,0) (y) {$j$};
        \node[ellipse, draw, align = left, text width = 1.8cm,
              minimum width = 2cm, 
              minimum height = 1.2cm] (c2) at (5,1.5) {Chain 2};
        \node[state] at (5.75,1.5) (s2) {$S_2$};      
        \node[below] at (6.25,1) {$P(S_2) = 0.05$};
        \node[below] at (5.75,2.75) {$\text{Cost} = 20 \text{ units}$};
        \node[ellipse, draw, align = left, text width = 1.8cm,
              minimum width = 2cm, 
              minimum height = 1.2cm] (c3) at (5,-1.5) {Chain 3};
        \node[below] at (6.25,-2) {$P(S_3) = 0.1$};
        \node[below] at (5.75,-0.25) {$\text{Cost} = 10 \text{ units}$};
        \node[state] at (5.75,-1.5) (s3) {$S_3$};

        \path[>=latex,
          auto=right,
          every loop]
            (x) edge node {$0.5$} (c1)
            (x) edge node {$0.5$} (y)
            (y) edge node {action $a$} (c2)
            (y) edge node {action $b$} (c3);
            
    \end{tikzpicture}
    \caption{The counterexample due to \cite{haviv1996constrained}. $i$ and $j$ are transient states. The decision maker can choose either action $a$ or action $b$ and correspondingly, the next state will be chain $2$ or chain $3$ respectively with probability $1$.}
    \label{fig:Haviv}
 \end{figure}
Formally, the problem can be stated in terms of \eqref{Eq.Optimize} and \eqref{Eq.Constraints} as follows.
\begin{subequations} \label{eq:Haviv}
    \begin{alignat}{3}
    &\min\limits_{\pi}   &\hspace{0.2cm} & V_{\pi}(i) := \mathds E_{\pi} \left[\sum_{t=0}^{\tau} c_t| x_0 = i\right], \label{eq:obj_Haviv}\\
    &\hspace{0.1cm}\text{s.t.}&\hspace{0.0001cm} & W_{\pi}(i) := \mathds E_{\pi} \left[\sum_{t=0}^{\tau} k_t| x_0 = i\right] \leq w \label{eq:safety_constraint},
    \end{alignat}
\end{subequations}
where $w = 0.125$. The safety function calculated from state $i$ is $W(i) = 0.5 \times 0.2 + 0.5 \times 0.05 = 0.125$ for action $a$ and $W(i) = 0.5 \times 0.2 + 0.5 \times 0.1 = 0.15$ for action $b$. The safety function calculated from state $j$ is $W(j) = 0.05$ for action $a$ and $W(j) = 0.1$ for action $b$. Therefore, action $b$ is feasible and optimal if we start from state $j$. However, action $b$ is infeasible and therefore not optimal if we start from state $i$. Thus \eqref{eq:Bellman_policy} does not hold in general for constrained multichain MDPs. This implies that a naive implementation of dynamic programming algorithms (and RL  algorithms based on Bellman equations) on constrained multichain MDPs may lead to solutions that are feasible but not optimal (i.e. conservative solutions). In the sequel, we seek to address this issue by using Zero-sum Stochastic (Markov) Game theory.  

\section{Proposed Solution}
In this section, we would like to find a solution to the problem defined above which satisfies Bellman's principle of optimality i.e. optimal policy $\pi^*$ which satisfies \eqref{eq:Bellman_policy}. 
Consider the constrained optimization problem defined by \eqref{Eq.Optimize} and \eqref{Eq.Constraints}. We shall now define the corresponding Lagrangian $\mathbf{L}_{\pi}$ as
\begin{align} \label{Lagrangian}
    \mathbf{L}_{\pi,  \Lambda} := \mathbf V_{\pi} + \Lambda \circ (\mathbf W_{\pi} - w \mathds{1}_{\mid E \mid}),
\end{align}
where $\Lambda = (\lambda)_{i \in E} \in \mathbb{R}^N$ represents the vector of Lagrange multipliers $\lambda$ with a multiplier associated to each statewise constraint. A constrained MDP can be considered a zero-sum Markov game between the state-dependent Lagrange multipliers $\lambda$ and the state-dependent policy $\pi$ with the Lagrangian in \eqref{Lagrangian} representing the cost for the decision maker (and the payoff for Lagrange multiplier) \cite{boyd2004convex}. For any policy $\pi$ we have,
\begin{equation}\label{eq:Lambda_eq}
\begin{aligned}
    \sup_{\lambda \geq 0} \mathbf{L}_{\pi, \Lambda} &= \sup_{\lambda \geq 0} \bigg(\mathbf V_{\pi} + \Lambda \circ (\mathbf W_{\pi} - w \mathds{1}_{\mid E \mid})\bigg) \\
    & = \begin{cases}\mathbf V_{\pi}, & \text {if } \mathbf W_{\pi} \leq w \mathds{1}_{\mid E \mid}, \\
                              \infty, & \text { otherwise. }\end{cases}
\end{aligned}
\end{equation}
Thus from \eqref{eq:Lambda_eq}, it can be concluded that if the value of the game does not exist then it means that the constrained optimization problem defined by \eqref{Eq.Optimize}, \eqref{Eq.Policy} and \eqref{Eq.Constraints} is infeasible. On the other hand, if the problem is feasible then the optimal value of $\Lambda$ should be $0$. We can therefore use Shapley's algorithm \cite{shapley1953stochastic} or its asynchronous variant (i.e. all the states are not updated simultaneously) for resolving the counterexample (fig. \ref{fig:Haviv}) due to \cite{haviv1996constrained}. Define the operator $\textbf{val}[\cdot]$ as the solution of the $ \sup_{\lambda \geq 0}\min_{\pi \in \Delta(A)}[\cdot]$ problem. Let the immediate Lagrangian cost be $d_t = c_t + \lambda_t(k_t - w)$, where $\lambda_t$ is the Lagrange multiplier associated with the state visited at time $t$. Suppose at time $t$, the algorithm has updated the Lagrangian values of states $1,\cdots,i$ then we can use the updated Lagrangian values of the same and use the old values for the remaining $i+1, \cdots, N$ states. Such a procedure is known as the Gauss-Seidel procedure and convergence of the same for zero-sum Markov games is proved in \cite{kushner2004gauss}. In the sequel, the superscript on $\mathbf{L}$ will denote the iteration number. The Gauss-Seidel procedure for the Markov game is the iteration in value space for successive substitutions. The substitutions are taken in the order $i = 1,2, \cdots, j, \cdots N$,
\begin{multline} \label{Eq.Gauss_Seidel}
     \mathbf{L}^{t}(i) = \sup_{\lambda_i \geq 0} \min_{\pi \in \Delta(A)} \bigg[ d_t + \sum\limits_{j=1}^{i-1} p_{ij}(\pi) L^{t}(j) \\ + \sum\limits_{j=i}^{N} p_{ij}(\pi) L^{t-1}(j) \bigg].
\end{multline} 
The following assumptions are required to be verified for the convergence of Algorithm \ref{shapley_alg} (based on the proof in \cite{kushner2004gauss} and references therein).
\begin{assumption} \label{assump:admissability}
There is at least one admissible policy $\Tilde{\pi}$ such that $\mathcal{B}_{\Tilde{\pi}}[\mathbf{L}_{\Tilde{\pi},\Lambda}]$ is a contraction with respect to the $\norm{\cdot}_{\infty}$, and the infimum of the costs over all admissible policies is bounded from below. Furthermore, $\mathcal{B}_{\pi}[\mathbf{L}_{\pi,\Lambda}]$ is a contraction  with respect to the $\norm{\cdot}_{\infty}$ for any admissible policy $\pi$ for which the associated cost is bounded.
\end{assumption}
\begin{assumption} \label{assump:continuity}
For every admissible policy $\pi$, $\mathbf{L}_{\pi,  \Lambda}$ is a continuous functions of $\pi,\Lambda$ for every $i,j\in E$. 
\end{assumption}
\begin{assumption} \label{assump:bounded}
If the cost associated with the use of the control actions $a_{1}, \ldots$, $a_{t}$ (picked in accordance with control policy $\pi$) in sequence, is bounded, then
$$
P\left(\pi\right) \stackrel{t}{\rightarrow} 0 \text{ as } t \to \infty
$$
\end{assumption} 
\begin{assumption} \label{assump:saddle_point}
    There exists an admissible control policy $\pi^{*}(\cdot)$ such that the following saddle point condition is satisfied,
    \begin{equation} \label{eq:saddle_point}
        \sup_{\lambda_i \geq 0} \min_{\pi \in \Delta(A)} \mathbf{L}_{\pi, \Lambda} = \min_{\pi \in \Delta(A)} \sup_{\lambda_i \geq 0} \mathbf{L}_{\pi, \Lambda}.
    \end{equation}
\end{assumption}

Assumptions \ref{assump:admissability} and \ref{assump:continuity} are satisfied in our case if there exists an admissible policy $\pi(\cdot \mid i)$ $\forall i \in E$ and the costs $c_t$ (and correspondingly for $\mathbf{V}_\pi$) are continuous and bounded for any admissible policy $\pi(\cdot \mid i)$. Note that since $k_t$ represents probabilities of reaching an unsafe set, they automatically satisfy the above assumptions. Assumption \ref{assump:bounded} is also satisfied in our case since the matrix $P$ is sub-stochastic for any policy $\pi$ as per \eqref{eq:P_pi}. Finally, Assumption \ref{assump:saddle_point} is also satisfied in our case since it is proved in \cite{wisniewski2021safe} (Theorem 1) that there is no duality gap which implies that Slater's condition is satisfied \cite{boyd2004convex} and therefore there exists a unique saddle point of the form \eqref{eq:saddle_point}.  
\begin{algorithm}
    \caption{Asynchronous Lagrangian iteration Algorithm}
    \label{shapley_alg}
    \begin{algorithmic}[1] 
    \State \textbf{Input}: Initial state $i$, tolerance parameter  $\epsilon$
    
    \State Initialize $\mathbf{L} = 0$ 
    
    \While{$\norm{\mathbf{L}^{t+1} - \mathbf{L}^{t}} \geq \epsilon$} 
    
    \State Update $L^t(i)$ \begin{multline*} L^{t}(i) \gets \mathbf{val}\Bigg[d_t + \sum\limits_{j=1}^{i-1} p_{ij}(\pi) L^{t}(j) \\ + \sum\limits_{j=i}^{N} p_{ij}(\pi) L^{t-1}(j) \Bigg]\end{multline*}
    
    \State $i \gets i + 1$
    
    \State $t \gets t + 1$
    
    \EndWhile
    
    \end{algorithmic}
\end{algorithm}
Once the optimal Lagrangian $\mathbf{L}_\pi^*$ has been calculated, the optimal policy can be obtained as follows
\begin{equation}
    \pi^* = \arg\min\limits_{\pi \in \Delta(A)} \mathbf{L}_\pi^*.
\end{equation}

\subsection{Resolution of counterexample due to \cite{haviv1996constrained}}

Consider the example described by fig. \ref{fig:Haviv} and \eqref{eq:Haviv}. Using \eqref{Lagrangian}, we rewrite \eqref{eq:Haviv} as the following unconstrained optimization problem 
\begin{equation} \label{eq:Game_lagrangian1}
    L_{\pi}(i) = \min\limits_{\pi \in \Delta(A)} \sup\limits_{\lambda \geq 0} V_\pi(i) + \lambda(i)(W_\pi(i) - w), 
\end{equation}
with $i$ being the initial state in fig. \ref{fig:Haviv} and $w = 0.125$. Applying algorithm \ref{shapley_alg} on \eqref{eq:Game_lagrangian1} for state $i$ and $t=1$ results in,
\begin{multline} \label{eq:Game_lagrangian_state_i}
    L_{\pi}(i) = \textbf{val}\bigg[ 0.5(\lambda(i)(0.5 \times 0.2 - 0.125)) \\ + 0.5(\textbf{val}\left[L_{\pi}(j)\right]) \bigg], 
\end{multline}
and for state $j$ results in, 
\begin{equation}\label{eq:Game_lagrangian_state_j}
\begin{aligned} 
    L_{\pi}(j) &= \textbf{val}\left[\begin{array}{cc}
         & 20 + \lambda(j)(0.05 - 0.125) \\
         & 10 + \lambda(j)(0.10 - 0.125)
    \end{array} \right], \\
    L_{\pi}(j) &= 10.
\end{aligned}
\end{equation}
Therefore, the optimal action at state $j$ is action $b$. Substituting \eqref{eq:Game_lagrangian_state_j} in \eqref{eq:Game_lagrangian_state_i} for iteration $t = 2$ gives us the optimal action at state $i$ to be $b$ as well. Thus, Bellman's 'principle of optimality' is not violated. 

\section{RL algorithm for Constrained MDP}

In this section, we propose an off-policy RL algorithm based on the Lagrangian iteration algorithm presented above that learns the Lagrangian for an MDP with an unknown transition probability matrix and unknown costs while ensuring safety constraint satisfaction.
We begin by rewriting \eqref{Lagrangian} as an unconstrained minimization problem with $\pi \in \Delta(A)$ as the decision variable as follows,
\begin{equation}\label{eq:Unconstrained_Indicator}
    \min_{\pi \in \Delta(A)} \mathbf V_{\pi} + \mathcal{I}(\mathbf W_{\pi} - w \mathds{1}_{\mid E \mid}),
\end{equation}
where $\mathcal{I}$ is the indicator function (see Notation). There are many methods to approximate and solve convex optimization problems of the type \eqref{eq:Unconstrained_Indicator} such as the Primal-Dual approach or the $\log$ barrier approach (see \cite{boyd2004convex}). In this work, we will approximately solve \eqref{eq:Unconstrained_Indicator} using the $\log$ barrier approach.
\begin{equation}\label{eq:Unconstrained_log}
    \min_{\pi \in \Delta(A)} \mathbf V_{\pi} + \bigg(\frac{-1}{l}\bigg)\log(w \mathds{1}_{\mid E \mid} - \mathbf W_{\pi}),
\end{equation}
where $l>0$ decides the accuracy of approximation with the approximation being more accurate for higher values of $l$. The state-wise components $\lambda(i)_{i \in E}$ of Lagrange multiplier vector $\Lambda$ are obtained as
\begin{equation}
    \lambda(i) = \frac{-1}{l(W_{\pi}(i)-w)}.
\end{equation}
In the sequel, we define $\phi$ as
\begin{equation}\label{eq:phi}
    \phi = -\log(w \mathds{1}_{\mid E \mid} - \mathbf W_{\pi}), 
\end{equation}
and the approximate Lagrangian $\Tilde{\mathbf{L}}_\pi$ is defined as 
\begin{multline}\label{eq:L_tilde}
    \Tilde{\mathbf{L}}_\pi = [\Tilde{L}_\pi(1), \cdots, \Tilde{L}_\pi(i), \cdots, \Tilde{L}_\pi(N)] \\
    = \mathbf V_{\pi} + \bigg(\frac{-1}{l}\bigg)\log(w \mathds{1}_{\mid E \mid} - \mathbf W_{\pi}).
\end{multline}
The goal of the following RL algorithm is to learn $\Tilde{\mathbf{L}}_\pi$ by learning the corresponding state-action $Q$-values.
\begin{algorithm}
    \caption{Off-policy Lagrangian Q-learning Algorithm}
    \label{lagrange_learning_alg}
    \begin{algorithmic}[1] 
    \State \textbf{Input}: Initial state $i$, tolerance parameter $\varepsilon$, Initial policy $\pi = \frac{1}{\mid A \mid}$
    
    \State Initialize ${Q(i,a)} = 0$ $\forall i\in E, a \in A$ 
    
    \State Initialize learning rate $\alpha_i = 1$ $\forall i \in E$ 
    
    \While{$\norm{\mathbf{\Tilde{L}}^{t} - \mathbf{\Tilde{L}}^{t-1}} \geq \varepsilon$} 

    \State Observe state $i$ and update $f^{t}_i = f^{t-1}_i + 1$

    \State Update learning rate $\alpha_i = \frac{1}{f^t_i}$

    \State Choose control action $a \sim \pi^{t-1}$
    
    \State Update $Q(i,a)$ as \begin{multline*} Q^{t}(i,a) \gets (1-\alpha_i)Q^{t-1}(i,a) + \alpha_i(d_t + \mathbf{val}[\Tilde{L}^{t-1}(j)]), \\ 
                            \text{where } \mathbf{val}[\Tilde{L}^{t-1}(j)] = \min_{b \in A}Q^{t-1}(j,b) \end{multline*}

    \State $f^{t}_{i,a} = f^{t-1}_{i,a} + \begin{cases} 1, & \text{if } a = \arg\min_{a \in A} Q^t(i,a) \\
                                                  0, & \text{otherwise   }\end{cases}$
    
    \State $\pi^{t}(a \mid i) = \frac{f^{t}_{i,a}}{f^{t}_i}$
    
    \State Update state $i$ to the next state
    
    \State $t \gets t + 1$
    
    \EndWhile
    
    \end{algorithmic}
\end{algorithm}
We construct randomized policy $\pi$ (since finding deterministic control policies is an $NP$-hard problem for constrained MDPs \cite{feinberg2000constrained}) using the approximate state-action occupation measure $f_{i,a}$ which counts every time action $a$ was optimal, i.e., $a = \arg\min_{a \in A}Q(i, a)$ whenever state $i$ was visited. Finally, we introduce the approximate state occupation measure $f_i$ which counts every time state $i$ was visited. In the following algorithm, the superscript over $f_{i}$, $f_{i,a}$, $\mathbf{\Tilde{L}}$, $Q$ and $\pi$ indicates the iteration number. The following lemma (based on \cite{kearns2002near} where a similar lemma is given for a discounted unconstrained MDP) gives finite time guarantees and bounds on $\varepsilon$ in Algorithm \ref{lagrange_learning_alg} for learning the approximate Lagrangian.
\begin{lemma} \label{Lemma}
    Consider an unknown constrained MDP defined as per \eqref{Eq.Optimize}, \eqref{Eq.Policy}, \eqref{Eq.Constraints}, \eqref{eq:P_pi} and \eqref{eq:k_pi}. Define $c_{M}$ and $\phi_{M}$ to be the maximum among all the costs in the unknown constrained MDP (without loss of generality) and let $T$ be some finite time. Furthermore, since for any policy $\pi$,
    \begin{align*}
        \sum^N_{j = 1} \sum_{a \in A} p_{ij}(a) \pi_{ia} < 1 \quad \forall i \in E,
    \end{align*}
    the stopping probability can be defined as
    \begin{align} \label{eq:p_tau}
        p_{\tau}(\pi) := 1 - \sum^N_{j = 1} \sum_{a \in A} p_{ij}(a) \pi_{ia} > 0. 
    \end{align}
    In the sequel, a superscript $t$ on $p^t_{\tau}$ indicates the stopping probability at time $t$. Let $\gamma:= 1 - \min\limits_{i \in E} p_{\tau}(\pi) $ be the maximum probability of continuation (or not stopping) among all the states and $\varepsilon>0$ be a constant. If
    \begin{equation}
        T \geq \frac{1}{1-\gamma}\log\bigg(\frac{c_{M} + (1/l)\phi_{M}}{\varepsilon(1-\gamma)}\bigg) \text{ than, } 
    \end{equation}
    \begin{equation}
        \Tilde{\mathbf{L}}^*_\pi - \Tilde{\mathbf{L}}^T_\pi \leq \varepsilon.
    \end{equation}
\end{lemma}
\begin{proof}
    Consider a sample path $\omega$ and an admissible policy $\pi$ then the expected optimal Lagrangian along this path will be,
    \begin{multline}
        \Tilde{\mathbf{L}}^*_\pi(\omega) = c_1 + \frac{1}{l}\phi_1 +  (1-p^1_{\tau})\bigg(c_2 + \frac{1}{l}\phi_2\bigg) + \\ (1-p^1_{\tau})(1-p^2_{\tau}) \bigg(c_3 + \frac{1}{l}\phi_3\bigg) + \cdots,
    \end{multline}
    where $\phi_t$ is defined as \eqref{eq:phi} with the subscript $t$ indicating the barrier cost of violating \eqref{Eq.Constraints} at time $t$. The expected optimal Lagrangian for any sample path $\omega$ will be bounded as follows 
    \begin{equation}\label{eq:proof}
    \begin{aligned}
        \Tilde{\mathbf{L}}^*_\pi(\omega) &\leq c_1 + \frac{1}{l}\phi_1 + \gamma \bigg(c_2 + \frac{1}{l}\phi_2\bigg) + \gamma^2 \bigg(c_3 + \frac{1}{l}\phi_3\bigg) + \cdots \\
                                         &\leq \sum^T_{t = 1} \gamma^{t-1}\bigg(c_t + \frac{1}{l}\phi_t\bigg) + \bigg(c_{M} + \frac{1}{l}\phi_{M}\bigg)\gamma^T \sum^\infty_{t = 0}\gamma^t \\
                                         &\leq \Tilde{\mathbf{L}}^T_\pi(\omega) + \bigg(c_{M} + \frac{1}{l}\phi_{M}\bigg)\gamma^T \frac{1}{1-\gamma}.
    \end{aligned}        
    \end{equation}
    
    We now bound the last term in equation \eqref{eq:proof} as 
    \begin{equation}\label{eq:proof2}
        \bigg(c_{M} + \frac{1}{l}\phi_{M}\bigg)\gamma^T \frac{1}{1-\gamma} \leq \varepsilon.
    \end{equation}
    Note that since $0<\gamma<1$, $\varepsilon \to 0$ as $T \to \infty$. Rearranging the terms in \eqref{eq:proof2},
    \begin{equation}\label{eq:proof3}
        \begin{aligned}
            \gamma^T &\leq \frac{\varepsilon(1-\gamma)}{c_{M} + (1/l)\phi_{M}} \\
            T\log\gamma &\leq \log\bigg(\frac{\varepsilon(1-\gamma)}{c_{M} + (1/l)\phi_{M}}\bigg) \\
            T(-\log\gamma) &\geq \log\bigg(\frac{c_{M} + (1/l)\phi_{M}}{\varepsilon(1-\gamma)}\bigg).        
        \end{aligned}
    \end{equation}
    Substituting 
    \begin{equation}
        \log\gamma \approx \gamma - 1 + \text{h.o.t},           
    \end{equation}
    in \eqref{eq:proof3} (and ignoring higher order terms since $0<\gamma<1$) completes the proof.
\end{proof}
Lemma \ref{Lemma} shows that the decision-maker will experience less overall cost if the process $X_t$ stops earlier. Since the barrier function repels the decision-maker from the unsafe set $\mathcal{U}$ at each iteration, the optimal policy for the decision-maker should ensure convergence to the target set $\mathcal{T}$. 

\section{Conclusion}
 It can be concluded from this work, that RL or any dynamic programming based approach cannot be trivially applied to Reach-Avoid problems in the case of MDPs as Bellman's principle of optimality may not hold. We resolve this issue by using the $\mathbf{val}[\cdot]$ operator defined in Section V. The $\mathbf{val}[\cdot]$ operator forces the decision maker to select policies that are independent of the initial state and only depend on the current state. This also makes sense intuitively for counterexample due to \cite{haviv1996constrained} as once the state trajectory reaches state $j$ in fig. \ref{fig:Haviv}, it does not make sense to consider $P(S_1)$ as there is no chance of going to Chain $1$ from $j$. Future work will involve calculating bounds on the number of times the constraint is violated during the learning phase and proposing an on-policy RL algorithm.

\section{Acknowledgements}
Financial support from the Poul Due Jensen Foundation (Grundfos Foundation) for this research is gratefully acknowledged. 




 
\bibliographystyle{IEEEtran}
\bibliography{mybibfile}
 
\end{document}